\documentclass[a4paper,11pt,headings=big,DIV=12]{scrartcl}
\pdfoutput=1
\usepackage{amsmath,amssymb}
\usepackage{color,xcolor}
\definecolor{blue}{rgb}{0,0,0.5}
\usepackage[colorlinks,linkcolor=blue,urlcolor=blue,citecolor=blue]{hyperref}
\usepackage{graphicx}
\addtokomafont{disposition}{\rmfamily\boldmath}
\usepackage{cite}
\usepackage{xspace}
\usepackage{relsize}
\usepackage{bm}
\usepackage[utf8]{inputenc}
\usepackage{tabularx,colortbl}

\allowdisplaybreaks

\begin{document}

\begin{center}
{\Large\bfseries \boldmath
\vspace*{1.5cm}
Status of the $B\to K^*\mu^+\mu^-$ anomaly after Moriond 2017
}\\[0.8 cm]
{\large{
Wolfgang Altmannshofer$^a$, Christoph Niehoff$^b$, Peter Stangl$^b$,  David M. Straub$^b$
}\\[0.5 cm]
\small
$^a$ \textit{Department of Physics, University of Cincinnati, Cincinnati, Ohio 45221, USA}
\\[0.2cm]
$^b$ \textit{Excellence Cluster Universe, Boltzmannstr.~2, 85748~Garching, Germany}
}
\\[0.5 cm]
\footnotesize
E-Mail: \texttt{%
\href{mailto:altmanwg@ucmail.uc.edu}{altmanwg@ucmail.uc.edu}
\href{mailto:christoph.niehoff@tum.de}{christoph.niehoff@tum.de}
\href{mailto:peter.stangl@tum.de}{peter.stangl@tum.de}
\href{mailto:david.straub@tum.de}{david.straub@tum.de}
}
\end{center}

\bigskip

\begin{abstract}\noindent
Motivated by recent results by the ATLAS and CMS collaborations on the angular distribution of the $B \to K^* \mu^+\mu^-$ decay, we perform a state-of-the-art analysis of rare $B$ meson decays based on the $b \to s \mu \mu$ transition. Using standard estimates of hadronic uncertainties, we confirm the presence of a sizable discrepancy between data and SM predictions. We do not find evidence for a $q^2$ or helicity dependence of the discrepancy. The data can be consistently described by new physics in the form of a four-fermion contact interaction $(\bar s \gamma_\alpha P_L b)(\bar \mu \gamma^\alpha \mu)$.
Assuming that the new physics affects decays with muons but not with electrons, we make predictions for a variety of theoretically clean observables
sensitive to violation of lepton flavour universality.
\end{abstract}

\section{Introduction}

The angular distribution of the decay $B\to K^*\mu^+\mu^-$ has been known to be
a key probe of physics beyond the Standard Model (SM) at the LHC already before its start (see e.g.~\cite{Kruger:1999xa,Kruger:2005ep,Bobeth:2008ij,Egede:2008uy,Altmannshofer:2008dz})
and the observable $S_5$ was recognized early on to be particularly promising
\cite{Altmannshofer:2008dz,Bharucha:2010bb}. A different normalization for
this observable, reducing form factor uncertainties, was suggested in
ref.~\cite{DescotesGenon:2012zf}, rebranded as $P_5'$.
While $B$ factory and Tevatron measurements of the forward-backward asymmetry and longitudinal
polarization fraction had been in agreement with SM expectations
\cite{Wei:2009zv,Aaltonen:2011ja,Lees:2015ymt},
in 2013, the LHCb collaboration announced the observation of a tension in the observable $P_5'$ at
the level of around three standard deviations.
It was quickly recognized~\cite{Descotes-Genon:2013wba}
that a new physics (NP) contribution to the Wilson
coefficient $C_9$ of a  semi-leptonic vector operator was able to explain this
``$B\to K^*\mu^+\mu^-$ anomaly'', confirmed few days later by an independent analysis
\cite{Altmannshofer:2013foa} and also by other groups with different methods
\cite{Beaujean:2013soa,Hurth:2013ssa}.
Further measurements have shown additional tensions, e.g.\ branching ratio
measurements in $B\to K\mu^+\mu^-$ and $B_s\to\phi \mu^+\mu^-$
\cite{Aaij:2014pli,Aaij:2015esa}, as well as, most notably, a hint for lepton
flavour non-universality in $B^+\to K^+\ell^+\ell^-$ decays~\cite{Aaij:2014ora}.
While progress has also been made on the theory side, most notably improved
$B\to K^*$ form factors from lattice QCD (LQCD)~\cite{Horgan:2013hoa,Horgan:2015vla}
and light-cone sum rules (LCSR)~\cite{Straub:2015ica},
the ``anomaly'' has also led to a renewed scrutiny of theoretical uncertainties due to form factors
\cite{Jager:2014rwa,Descotes-Genon:2014uoa,Capdevila:2017ert}
as well as non-factorizable  hadronic effects
\cite{Lyon:2014hpa,Ciuchini:2015qxb,Chobanova:2017ghn}
(cf. also the earlier works
\cite{Khodjamirian:2010vf,Beylich:2011aq,Khodjamirian:2012rm,Jager:2012uw}).

In 2015, the LHCb collaboration presented their $B\to  K^*\mu^+\mu^-$ angular
analysis based on the full Run~1 data set, confirming the tension found earlier
\cite{Aaij:2015oid}.
Several updated global analyses have confirmed that a consistent description
of the tensions in terms of NP is possible
\cite{Altmannshofer:2014rta,Descotes-Genon:2015uva,Hurth:2016fbr},
while an explanation in terms of an unexpectedly large hadronic effect cannot
be excluded.
Recent analyses by Belle~\cite{Abdesselam:2016llu,Wehle:2016yoi} also seem to indicate tensions
in angular observables consistent with LHCb.
At Moriond Electroweak 2017, ATLAS~\cite{ATLAS-CONF-2017-023} and CMS~\cite{CMS-PAS-BPH-15-008} finally presented their preliminary
results for the angular observables based on the full Run~1 data sets.
The aim of the present paper is to reconsider the status of the
$B\to K^*\mu^+\mu^-$ anomaly in view of these results.
Our analysis is built on our previous global analyses of NP in $b\to s$
transitions
\cite{Altmannshofer:2011gn,Altmannshofer:2012az,Altmannshofer:2013foa,Altmannshofer:2014rta}
and makes use of the open source code
\texttt{flavio}~\cite{flavio}.

\section{Effective Hamiltonian and observables}\label{sec:Heff}

The effective Hamiltonian for $b\to s$ transitions can be written as
\begin{equation}
\label{eq:Heff}
\mathcal{H}_\text{eff} = - \frac{4\,G_F}{\sqrt{2}} V_{tb}V_{ts}^* \frac{e^2}{16\pi^2}
\sum_i
(C_i O_i + C'_i O'_i) + \text{h.c.}
\end{equation}
and we consider NP effects in the following set of dimension-6 operators,
\begin{align}
O_9 &=
(\bar{s} \gamma_{\mu} P_{L} b)(\bar{\ell} \gamma^\mu \ell)\,,
&
O_9^{\prime} &=
(\bar{s} \gamma_{\mu} P_{R} b)(\bar{\ell} \gamma^\mu \ell)\,,\label{eq:O9}
\\
O_{10} &=
(\bar{s} \gamma_{\mu} P_{L} b)( \bar{\ell} \gamma^\mu \gamma_5 \ell)\,,
&
O_{10}^{\prime} &=
(\bar{s} \gamma_{\mu} P_{R} b)( \bar{\ell} \gamma^\mu \gamma_5 \ell)\,.\label{eq:O10}
\end{align}
We neither consider new physics in scalar operators, as they are strongly constrained
by ${B_s\to\mu^+\mu^-}$ (see~\cite{Altmannshofer:2017wqy} for a recent analysis),
nor in dipole operators, which are strongly constrained by inclusive and exclusive radiative
decays (see~\cite{Paul:2016urs} for a recent analysis).
We also do not consider new physics in four-quark operators, although
an effect in certain $b\to c\bar cs$ operators could potentially relax some of
the tensions in $B\to K^*\mu^+\mu^-$ angular observables \cite{Jager:2017gal}.

In our numerical analysis, we include the following observables.
\begin{itemize}
\item Angular observables in $B^0\to K^{*0}\mu^+\mu^-$ measured by
CDF~\cite{CDFupdate}, LHCb~\cite{Aaij:2015oid}, ATLAS*~\cite{ATLAS-CONF-2017-023}, and CMS*~\cite{Khachatryan:2015isa,1385600,CMS-PAS-BPH-15-008},
\item $B^{0,\pm}\to K^{*0,\pm}\mu^+\mu^-$ branching ratios by LHCb*~\cite{Aaij:2014pli,Aaij:2016flj},
CMS ~\cite{Khachatryan:2015isa,1385600},
and CDF~\cite{CDFupdate},
\item $B^{0,\pm}\to K^{0,\pm}\mu^+\mu^-$ branching ratios by LHCb~\cite{Aaij:2014pli} and CDF~\cite{CDFupdate},
\item $B_s\to\phi\mu^+\mu^-$ branching ratio by LHCb*~\cite{Aaij:2015esa} and CDF~\cite{CDFupdate},
\item $B_s\to\phi\mu^+\mu^-$ angular observables by LHCb*~\cite{Aaij:2015esa},
\item the branching ratio of the inclusive decay $B\to X_s\mu^+\mu^-$ measured
by BaBar~\cite{Lees:2013nxa}.
\end{itemize}
Items marked with an asterisk have been updated since our previous global fit
\cite{Altmannshofer:2014rta}.
Concerning  $B^0\to K^{*0}\mu^+\mu^-$, both LHCb and ATLAS have performed
measurements of CP-averaged angular
observables $S_i$ as well as of the closely related ``optimized''
observables $P_i'$. While LHCb gives also the full correlation matrices and the
choice of basis is thus irrelevant (up to non-Gaussian effects which are anyway
impossible to take into account using publicly available information), ATLAS
does not give correlations, so the choice can make a difference in principle.
We have chosen to use the $P_i'$ measurements, but have explicitly checked that
the best-fit regions and pulls do not change significantly when using the $S_i$
observables.

We do \textit{not} include the following measurements.
\begin{itemize}
 \item Angular observables in $B\to K\mu^+\mu^-$, which are only relevant
 in the presence of scalar or tensor operators~\cite{Beaujean:2015gba},
 \item measurements of lepton-averaged observables, as  we want to focus
 on new physics in $b\to s\mu^+\mu^-$ transitions,
 \item the Belle measurement of $B\to K^*\mu^+\mu^-$ angular observables
~\cite{Wehle:2016yoi},
 as it contains an unknown mixture of $B^0$ and $B^\pm$ decays that receive
 different non-factorizable corrections at low $q^2$,
 \item the LHCb measurement of the decay $\Lambda_b\to\Lambda\mu^+\mu^-$~\cite{Aaij:2015xza},
 as it still suffers from large experimental uncertainties and the central
 values of the measurement are not compatible with any viable short-distance
 hypothesis~\cite{Meinel:2016grj}.
\end{itemize}
We do not make use of the LHCb analysis attempting to separately extract
the short- and long-distance contributions to the $B^+\to K^+\mu^+\mu^-$
decay~\cite{Aaij:2016cbx}, but we note that these results are in qualitative
agreement with our estimates of long-distance contributions to this decay.
Finally, we do not include the decay $B_s\to\mu^+\mu^-$ in our fit, as it can
be affected by scalar operators, as discussed above.

For all these semi-leptonic observables, that are measured in bins of $q^2$,
we discard the following bins from our numerical analysis.
\begin{itemize}
\item Bins below the $J/\psi$ resonance that extend above 6~GeV$^2$. In this region,
 theoretical calculations based on QCD factorization are not reliable~\cite{Beneke:2001at}.
\item Bins above the $\psi(2S)$ resonance that are less than 4~GeV$^2$ wide.
This is because theoretical predictions are only valid for sufficiently global,
i.e.\ $q^2$-integrated, observables in this region~\cite{Beylich:2011aq}.
\item Bins with upper boundary \textit{at or below} 1~GeV$^2$, because this region
is dominated by the photon pole and thus by dipole operators, while we are
interested in the effect of semi-leptonic operators in this work.
\end{itemize}

For the SM predictions of these observables, we refer the reader to
refs.~\cite{Altmannshofer:2014rta,Straub:2015ica}, where the calculations,
inputs, and parametrization of hadronic uncertainties have been discussed in
detail. Our predictions are based on the implementation of these calculations
in the open source code \texttt{flavio}~\cite{flavio}.
With respect to our previous analysis~\cite{Altmannshofer:2014rta}, we use
improved predictions for $B \to K^*$ and $B_s \to \phi$ form factors
from~\cite{Straub:2015ica} and $B \to K$ form factors from~\cite{Bailey:2015dka}.
Note that the $B \to K$ form factors from~\cite{Bailey:2015dka} have substantially smaller
uncertainties compared to the ones used in~\cite{Altmannshofer:2014rta} which
were based on the results in~\cite{Ball:2004ye,Bartsch:2009qp,Bouchard:2013pna}.
The increased tension due to these form factors was also pointed out in \cite{Du:2015tda}.

\section{Results and discussion}

From the measurements and theory predictions, we construct a $\chi^2$ function
where theory uncertainties are combined with experimental uncertainties,
such that the $\chi^2$ only depends on the Wilson coefficients. Both for the
theoretical and the experimental uncertainties, we take into account all known
correlations and approximate the uncertainties as (multivariate) Gaussians,
and we neglect the dependence of the uncertainties on the NP contributions.
This procedure, which was proposed in~\cite{Altmannshofer:2014rta} and later
adopted by other groups
\cite{Descotes-Genon:2015uva}
is  implemented in \texttt{flavio} as the \texttt{FastFit} class.

From the observable selection discussed in section~\ref{sec:Heff}, we end up
with a total number of 86 measurements of 81 distinct observables.
These observables are not independent, but their theoretical and experimental uncertainties
are correlated. We take into account the experimental correlations where known
(this is the case only for the angular analyses
of $B\to K^*\mu^+\mu^-$ and $B_s\to\phi\mu^+\mu^-$ by LHCb), and
include all theory correlations.
Before considering NP effects, we can evaluate
the $\chi^2$ function within the SM
to get a feeling of the agreement of the data with the SM hypothesis. However, this
absolute $\chi^2$ is not uniquely defined.
For instance, averaging multiple measurements of
identical observables by different experiments before they enter the $\chi^2$,
we obtain $\chi^2_\text{SM}=98.5$ for 81 observables.
Adding all individual measurements separately instead, we obtain
$\chi^2_\text{SM}=100.6$ for 86 measurements. For the $\Delta\chi^2$ used in the
remainder of the analysis, these procedures are equivalent.

\subsection{New physics in individual Wilson coefficients} \label{sec:1d}

As a first step, we switch on NP contributions in individual Wilson coefficients,
determine the best-fit point in the one- or
two-dimensional space, and evaluate the $\chi^2$ difference $\Delta\chi^2$ with
respect to the SM point. The ``pull'' in $\sigma$
is then defined as $\sqrt{\Delta\chi^2}$ in the one-dimensional case,
while in the two-dimensional case it can be evaluated using the inverse cumulative
distribution function of the $\chi^2$ distribution with two degrees of freedom;
for instance, $\Delta\chi^2\approx 2.3$ for $1\sigma$.

The results are shown in table~\ref{tab:pulls}. We make the following observations.
\begin{itemize}
 \item The strongest pull is obtained in the scenario with NP in $C_9$ only
 and it amounts to slightly more than five standard deviations.
 Consistently with fits before the
 updated ATLAS and CMS measurements, the best-fit point corresponds to
 a value around $C_9\sim -1$, i.e. destructive interference with the SM Wilson
 coefficient. The increase in the significance for a non-standard $C_9$
 ($3.9\sigma$ in~\cite{Altmannshofer:2014rta} vs. $5.2\sigma$ here) can be
 largely traced back to the new and more precise form factors we are using,
 with only a moderate impact of the added experimental measurements.
 \item A scenario with NP in $C_{10}$ only also gives an improved fit, although
 less significantly than the $C_9$ scenario. We note that this suppression
 of $C_{10}$ by roughly 20\% would imply a suppression of the $B_s\to\mu^+\mu^-$
 branching ratio -- which, we stress again, we have not included in the fit --
 by roughly 35\%.
 \item A scenario with $C_9^\text{NP}=-C_{10}^\text{NP}$, that is well motivated
 by models with mediators coupling only to left-handed leptons, leads to a
 comparably good fit as the $C_9$-only scenario.
\end{itemize}

\renewcommand{\arraystretch}{1.5}
\begin{table}[tb]
\begin{center}
\begin{tabularx}{\textwidth}{ccccX}
\hline\hline
 ~~~~~~ Coeff. ~~~~~~ & ~~~~~~~ best fit ~~~~~~~ & ~~~~~~~~~~ $1\sigma$ ~~~~~~~~~~ & ~~~~~~~~~~ $2\sigma$ ~~~~~~~~~~ & pull \\
\hline\hline
\rowcolor[gray]{.9} $C_9^\text{NP}                  $ & $-1.21$ & [$-1.41$, $-1.00$] & [$-1.61$, $-0.77$] & $5.2\sigma$\\
                    $C_9^\prime                     $ & $+0.19$ & [$-0.01$, $+0.40$] & [$-0.22$, $+0.60$] & $0.9\sigma$\\
\rowcolor[gray]{.9} $C_{10}^\text{NP}               $ & $+0.79$ & [$+0.55$, $+1.05$] & [$+0.32$, $+1.31$] & $3.4\sigma$\\
                    $C_{10}^\prime                  $ & $-0.10$ & [$-0.26$, $+0.07$] & [$-0.42$, $+0.24$] & $0.6\sigma$\\
\rowcolor[gray]{.9} $C_9^\text{NP}=C_{10}^\text{NP} $ & $-0.30$ & [$-0.50$, $-0.08$] & [$-0.69$, $+0.18$] & $1.3\sigma$\\
                    $C_9^\text{NP}=-C_{10}^\text{NP}$ & $-0.67$ & [$-0.83$, $-0.52$] & [$-0.99$, $-0.38$] & $4.8\sigma$\\
\rowcolor[gray]{.9} $C_9^\prime=C_{10}^\prime       $ & $+0.06$ & [$-0.18$, $+0.30$] & [$-0.42$, $+0.55$] & $0.3\sigma$\\
                    $C_9^\prime=-C_{10}^\prime      $ & $+0.08$ & [$-0.02$, $+0.18$] & [$-0.12$, $+0.28$] & $0.8\sigma$\\
\hline
\rowcolor[gray]{.9} $C_9^\text{NP},\ C_{10}^\text{NP}$ & ($-1.15$, $+0.26$) &  ---   &        ---         & $5.0\sigma$\\
                    $C_9^\text{NP},\ C_{9}^\prime    $ & ($-1.25$, $+0.59$) &  ---   &        ---         & $5.3\sigma$\\
\rowcolor[gray]{.9} $C_9^\text{NP},\ C_{10}^\prime   $ & ($-1.34$, $-0.39$) &  ---   &        ---         & $5.4\sigma$\\
                    $C_9^\prime,\ C_{10}^\text{NP}   $ & ($+0.25$, $+0.83$) &  ---   &        ---         & $3.2\sigma$\\
\rowcolor[gray]{.9} $C_9^\prime,\ C_{10}^\prime      $ & ($+0.23$, $+0.04$) &  ---   &        ---         & $0.5\sigma$\\
                    $C_{10}^\text{NP},\ C_{10}^\prime$ & ($+0.79$, $-0.05$) &  ---   &        ---         & $3.0\sigma$\\
\hline\hline
\end{tabularx}
\end{center}
\caption{Best-fit values and pulls in sigma between the best-fit point and the
SM point for scenarios with NP in one or two Wilson coefficients. For the
one-dimensional cases, we also show the 1 and $2\sigma$ best-fit ranges. For
two of the two-dimensional cases, the best-fit regions are shown in
fig~\ref{fig:2dplots}.}
\label{tab:pulls}
\end{table}

To understand where the large global tension comes from, it is instructive to
perform one-dimensional fits with NP in $C_9$ using only a subset of the data.
We find for instance that
\begin{itemize}
 \item measurements of the $B_s\to\phi\mu^+\mu^-$ branching ratio alone lead
 to a pull of $3.5\sigma$,
 \item all branching ratio measurements combined lead to a pull of $4.6\sigma$,
 \item the $B\to K^*\mu^+\mu^-$ angular analysis by LHCb alone leads to a pull
 of $3.0\sigma$,
 \item the new $B\to K^*\mu^+\mu^-$ angular analysis by CMS reduces the pull,
 but the new ATLAS measurement increases it.
\end{itemize}

The significance of the tension between the branching ratio measurements and the
corresponding SM predictions depends strongly on the form factors used.
To estimate the possible impact of underestimated form factor uncertainties,
we repeat the fit with NP in $C_9$, doubling the form factor uncertainties
with respect to our nominal fit.
We find that the pull is reduced from $5.2\sigma$ to $4.0\sigma$. Significant
tensions remain in this scenario, indicating that underestimated form factor
uncertainties are likely not the only source of the discrepancies.

We also perform a fit doubling the uncertainties of the non-factorizable
hadronic corrections (see~\cite{Altmannshofer:2014rta} for details on how we
estimate these uncertainties). We find a reduced pull of $4.4\sigma$.

\subsection{New physics in pairs of Wilson coefficients} \label{sec:2d}

Next, we consider pairs of Wilson coefficients. In the last four rows of
table~\ref{tab:pulls}, we show the best-fit points and pulls for four different
scenarios. We observe that adding one of the primed coefficients does not
improve the fit substantially.

\begin{figure}[tbp]
\centering
\includegraphics[width=0.5\textwidth]{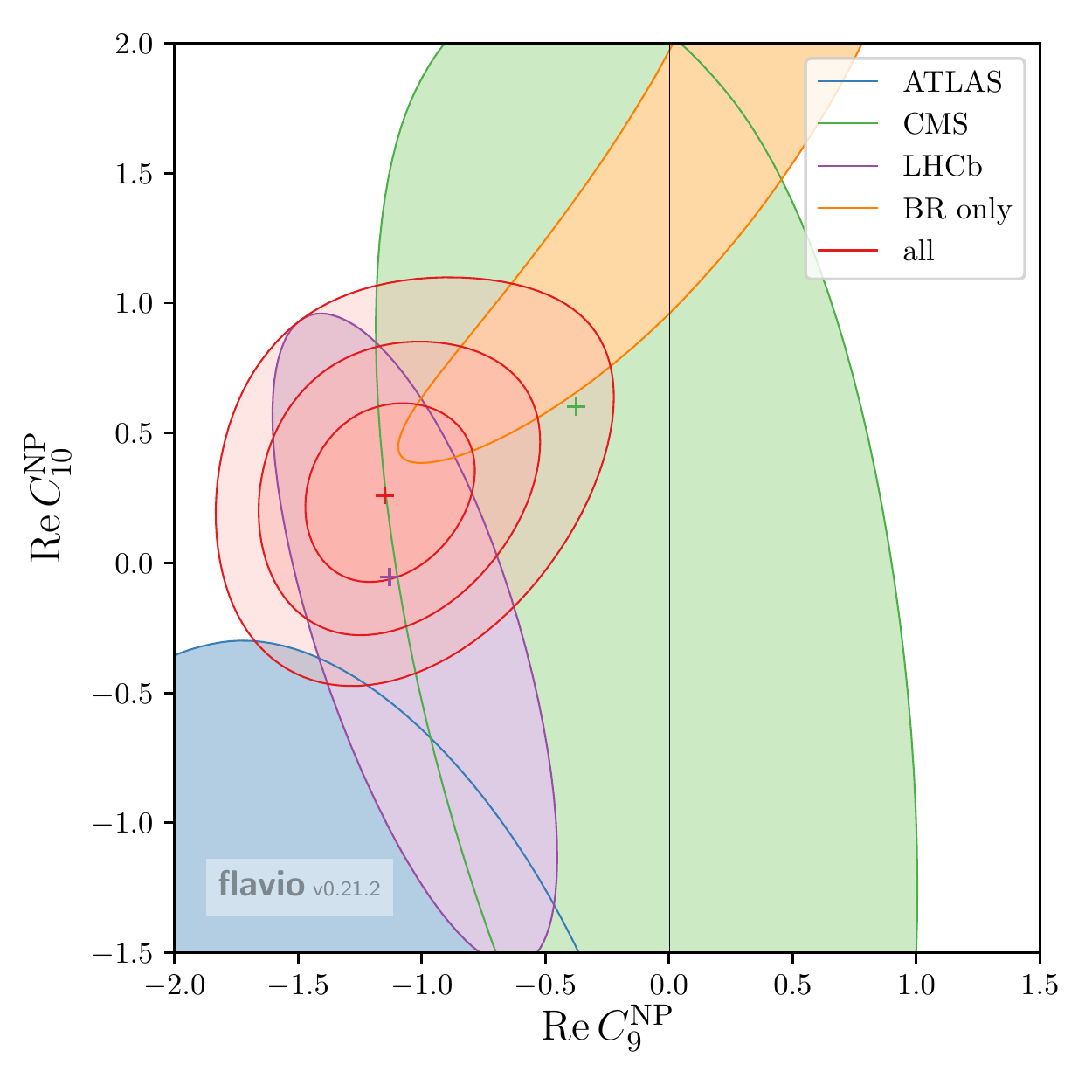}%
\includegraphics[width=0.5\textwidth]{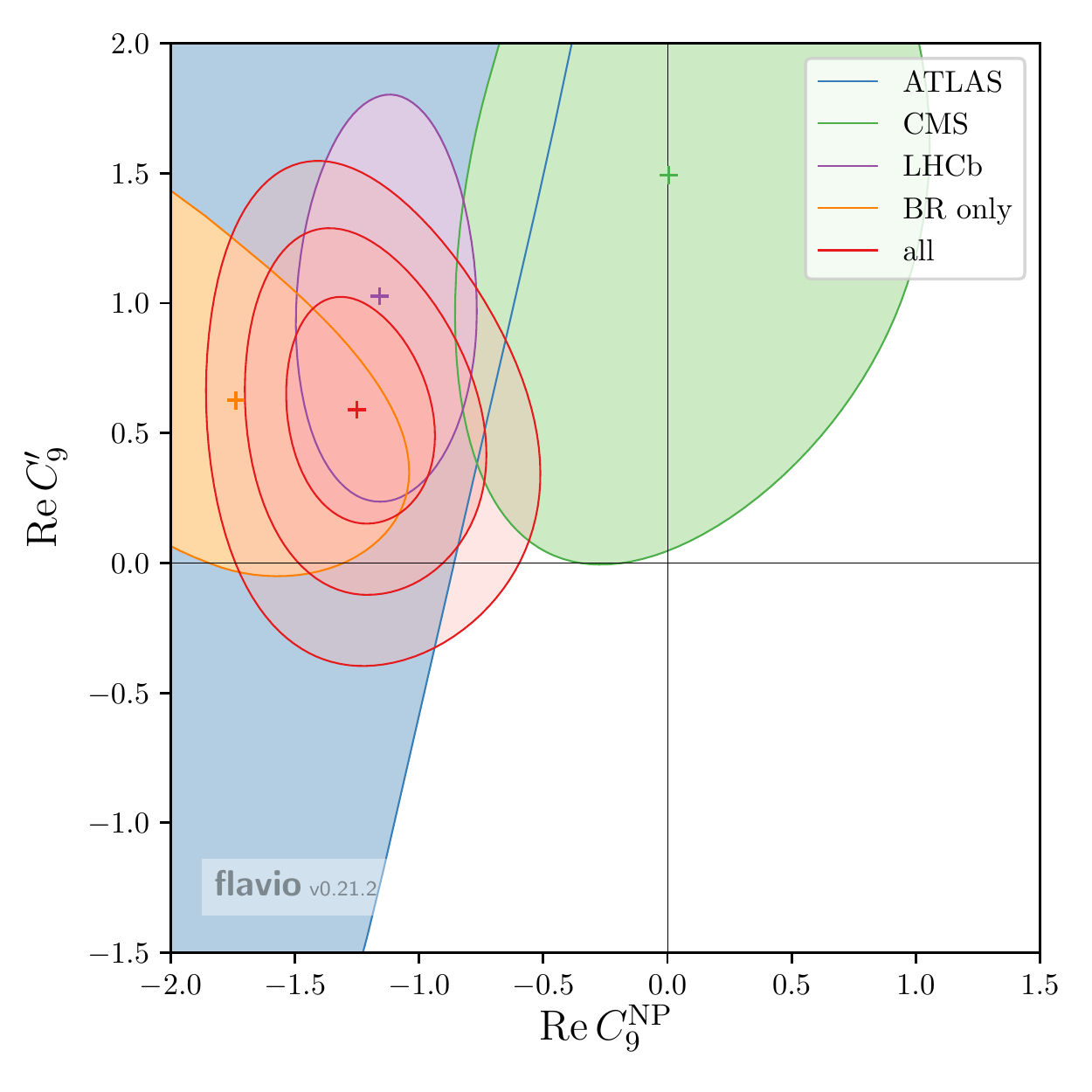}
\caption{Two-dimensional constraints in the plane of NP contributions to
the real parts of the Wilson coefficients $C_9$ and $C_{10}$ (left) or
$C_9$ and $C_9'$ (right), assuming all other Wilson coefficients to be SM-like.
For the constraints from the $B\to K^*\mu^+\mu^-$ and $B_s\to\phi\mu^+\mu^-$
angular observables from individual experiments as well as for the constraints
from branching ratio measurements of all experiments (``BR only''), we show the
$1\sigma$ ($\Delta\chi^2\approx 2.3$) contours, while for the global fit (``all''),
we show the 1, 2, and $3\sigma$ contours.}
\label{fig:2dplots}
\end{figure}

In fig.~\ref{fig:2dplots} we plot contours of constant $\Delta\chi^2$ in the planes of two Wilson
coefficients for the scenarios with NP in $C_9$ and $C_{10}$ or in $C_9$ and $C_9'$,
assuming the remaining coefficients to be SM-like.
In both plots, we show the 1, 2, and $3\sigma$ contours for the global fit,
but also $1\sigma$ contours showing the constraints coming from the
angular analyses of individual experiments, as well as from branching ratio
measurements of all experiments.

We observe that the individual constraints are all compatible with the global
fit at the $1\sigma$ or $2\sigma$ level. While the CMS angular analysis shows
good agreement with the SM expectations, all other individual constraints show
a deviation from the SM. In view of their precision, the angular analysis and
branching ratio measurements of LHCb still dominate the global fit
(cf.~Figs.~\ref{fig:pred:br},~\ref{fig:pred:ks-p},~\ref{fig:pred:ks-s}
and~\ref{fig:pred:phi-s}), leading
to a similar allowed region as in previous analyses. We do not find any
significant preference for non-zero NP contributions in $C_{10}$ or $C_9'$ in
these two simple scenarios.

\begin{figure}[tbp]
\centering
\includegraphics[width=0.5\textwidth]{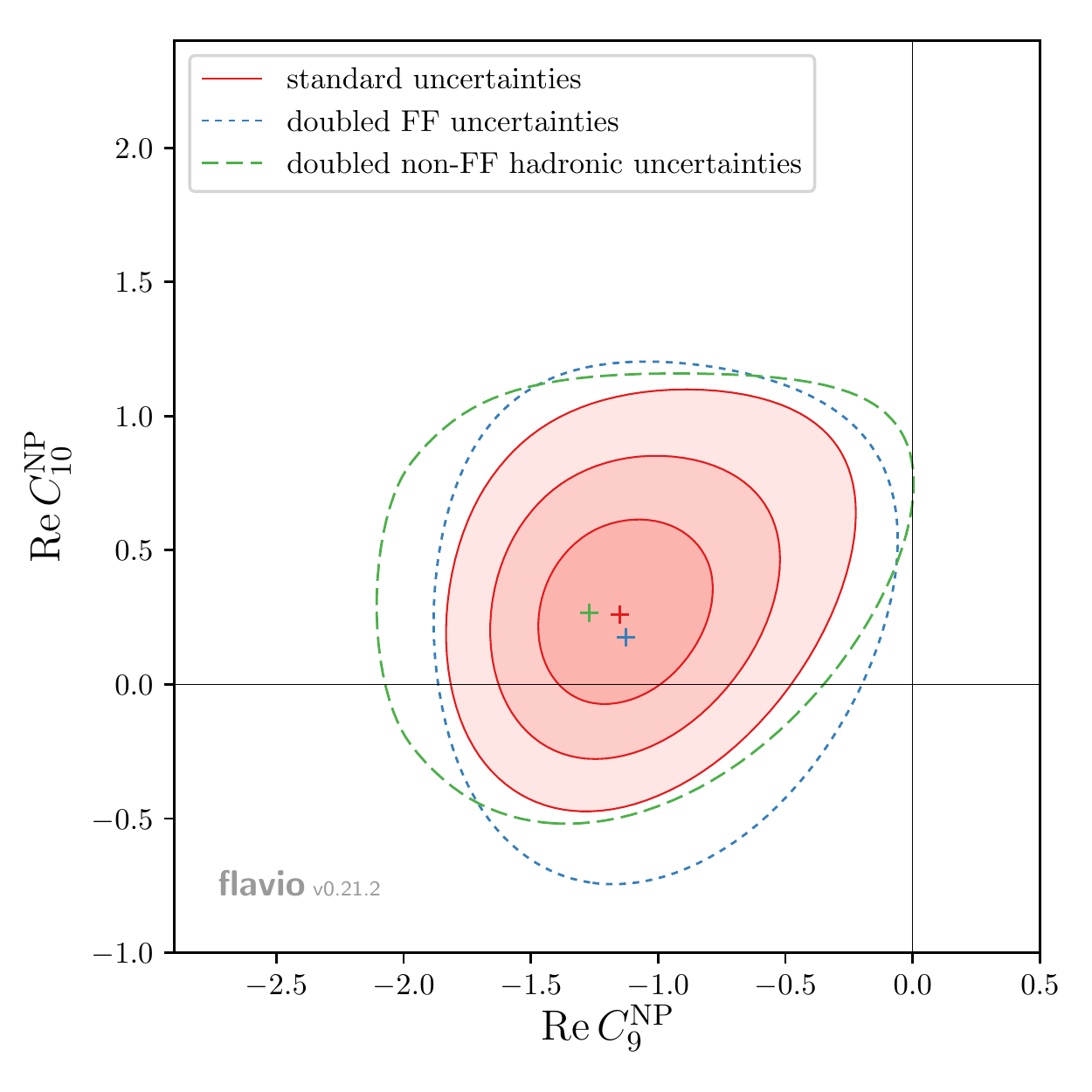}%
\includegraphics[width=0.5\textwidth]{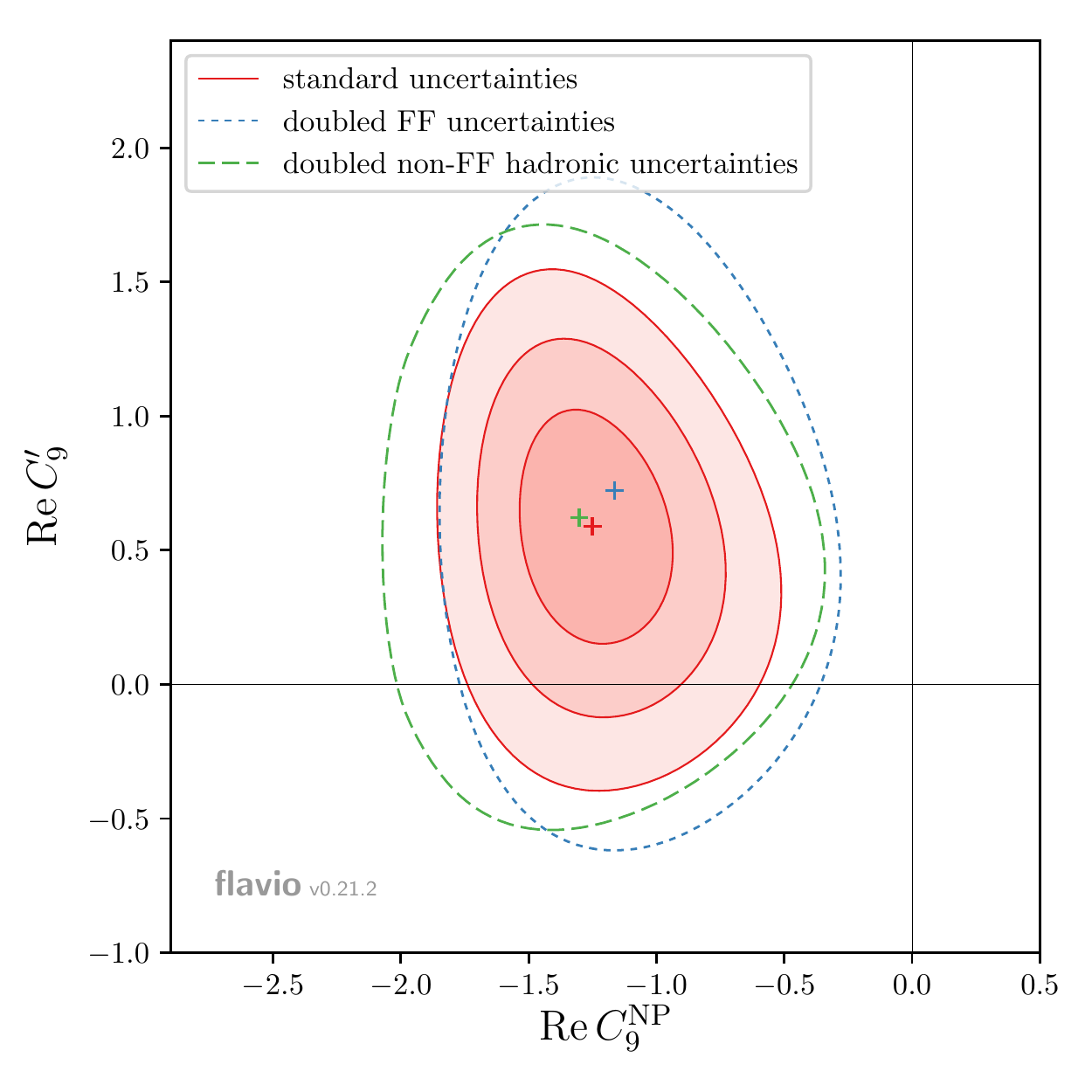}
\caption{Allowed regions in the Re$(C_9^\text{NP})$-Re$(C_{10}^\text{NP})$ plane (left) and the Re$(C_9^\text{NP})$-Re$(C_9^\prime)$ plane (right). In red the $1\sigma$, $2\sigma$, and $3\sigma$ best fit regions with nominal hadronic uncertainties. The green dashed and blue short-dashed contours correspond to the $3\sigma$ regions in scenarios with doubled uncertainties from non-factorizable corrections and doubled form factor uncertainties, respectively.}
\label{fig:uncertainties}
\end{figure}

Similarly to our analysis of scenarios with NP in one Wilson coefficient,
we repeat the fits doubling the form factor uncertainties and doubling the
uncertainties of non-factorizable corrections.
For NP in $C_9$ and $C_{10}$, we find that the pull is reduced from $5.0\sigma$
to $3.7\sigma$ and $4.1\sigma$, respectively.
For NP in $C_9$ and $C_9^\prime$ the pull is reduced from
$5.3\sigma$ to $4.1\sigma$ and $4.4\sigma$,
respectively. The impact of the inflated uncertainties is also
illustrated in Fig.~\ref{fig:uncertainties}. Doubling the
hadronic uncertainties is not sufficient to achieve agreement
between data and SM predictions at the $3\sigma$ level.

\subsection{New physics or hadronic effects?}

\begin{figure}[tbp]
\includegraphics[width=\textwidth]{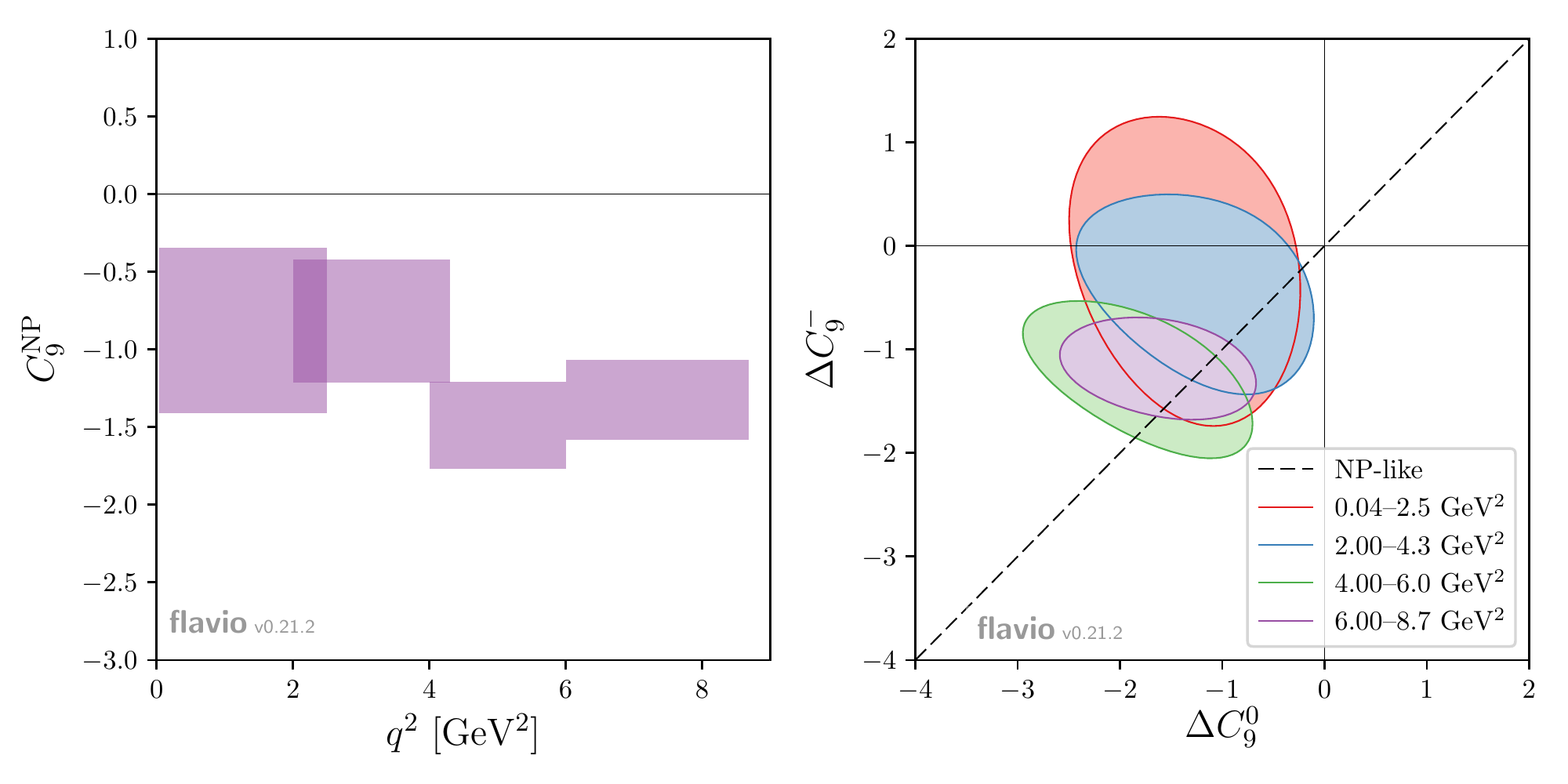}
\caption{Left: preferred $1\sigma$ ranges for a new physics contribution to $C_9$ from fits in different $q^2$ bins.
Right: preferred $1\sigma$ ranges for helicity dependent contributions to $C_9$ from fits in different $q^2$ bins. The dashed diagonal line corresponds to a helicity universal contribution, as predicted by new physics.}
\label{fig:deltac9}
\end{figure}

It is conceivable that hadronic effects that are largely underestimated could mimic new physics in the Wilson coefficient $C_9$~\cite{Lyon:2014hpa}.
As first quantified in~\cite{Altmannshofer:2015sma}
and later considered in~\cite{Descotes-Genon:2015uva,Ciuchini:2015qxb,Chobanova:2017ghn,Capdevila:2017ert}, there are ways to test this possibility by studying the $q^2$ and helicity dependence of a non-standard effect in $C_9$.

Without loss of generality, any photon-mediated hadronic contribution to the $B \to K^* \mu^+\mu^-$
helicity amplitudes can be expressed as a $q^2$ and helicity dependent shift in $C_9$, since
the photon has a vector-like coupling to leptons and flavour-violation always
involves left-handed quarks in the SM.
A new physics contribution to the Wilson coefficient $C_9$ is by definition independent of the di-muon invariant mass $q^2$, and it is universal for all three helicity amplitudes.
For hadronic effects, the situation is rather different.
It can be argued that hadronic effects in the $\lambda = +$ helicity amplitudes are suppressed~\cite{Jager:2012uw} and a priori there is no reason to expect that hadronic effects in the $\lambda=0$ and $\lambda = -$ amplitudes are of the same size.
Moreover, one would naively expect that hadronic effects that can arise e.g. from charm loops show a non-trivial $q^2$ behaviour. However, we would like to stress that no robust predictions about the precise properties of the hadronic effects can be made at present.

Another interesting possibility is to have NP contributions in $b\to c\bar cs$ operators
as speculated in \cite{Lyon:2014hpa} and recently worked out in \cite{Jager:2017gal}.
In this case, the shift in $C_9$ would be $q^2$ dependent, but helicity \textit{independent}
up to corrections of order $\alpha_s$ and $\Lambda_\text{QCD}/m_b$.

In order to understand if the data shows preference for a non-trivial $q^2$ dependence, we perform a series of fits to non-standard contributions to the Wilson coefficient $C_9$ in individual bins of $q^2$, using
$B^0\to K^{*0}\mu^+\mu^-$ measurements only.
In particular, we consider separately the experimental data in bins below 2.5~GeV$^2$, between 2~GeV$^2$ and 4.3~GeV$^2$, between 4~GeV$^2$ and 6~GeV$^2$, and between 6~GeV$^2$ and 8.7~GeV$^2$ (the overlaps are due to the
different binning unfortunately still used by different experiments).
While the latter bin is not included in our NP fit as discussed in section~\ref{sec:Heff},
we include it here as we are explicitly interested in the hadronic effects mimicking
 a shift in $C_9$.
The results are shown in the left plot of Fig.~\ref{fig:deltac9}.
While the significance of the tension is more pronounced in the region above 4~GeV$^2$,
this is not surprising as the observables are more sensitive to $C_9$ in this region.
At $1\sigma$, the fits are compatible with a flat $q^2$ dependence.
Moreover, every single bin shows a preference for a shift in $C_9$,
compatible with a constant new physics contribution of $C_9^\text{NP} \sim -1$.

In the right plot of Fig.~\ref{fig:deltac9} we show results of fits that allow for helicity dependent shifts in the Wilson coefficient $C_9$, which we denote as $\Delta C_9^0$ and $\Delta C_9^-$. As before we split the data into $q^2$ bins. The fit results are perfectly consistent with a universal effect $\Delta C_9^0 = \Delta C_9^-$ for each individual $q^2$ bin. Furthermore, we also find that the fit results of the different $q^2$ bins are consistent with each other.

The absence of a $q^2$ and helicity dependence is intriguing, but cannot exclude a hadronic effect as the origin of the apparent discrepancies.

\subsection{Predictions for LFU Observables}

As discussed, the ``$B \to K^* \mu^+\mu^-$ anomaly'' can be consistently described by new physics contributions to Wilson coefficients of the effective Hamiltonian~(\ref{eq:Heff}).
In order to determine the best-fit values for the various Wilson coefficients, we considered exclusively data on rare decays with muons in the final state. In this section, we use the obtained best-fit ranges from sections~\ref{sec:1d} and~\ref{sec:2d} to make predictions for theoretically clean lepton flavour universality (LFU) observables.

In contrast to hadronic effects, NP can lead to lepton flavour non-universality.
NP predictions for LFU observables depend on additional assumptions how the NP affects $b \to s e e$ transitions.
Well motivated are NP scenarios where $b \to s e e$ transitions remain approximately SM like. This is realized for example in models that are based on the $L_\mu - L_\tau$ gauge symmetry~\cite{Altmannshofer:2014cfa,Altmannshofer:2015mqa} and is also naturally the case in models based on partial compositeness~\cite{Niehoff:2015bfa}.
We will therefore assume that $b \to s e e$ transitions are unaffected by NP. We use our fit results to map out the allowed ranges for a variety of LFU observables.

We consider the following ratios of branching ratios~\cite{Hiller:2003js,Hiller:2014ula}
\begin{equation}
R_K = \frac{\text{Br}(B \to K \mu^+\mu^-)}{\text{Br}(B \to K e^+e^-)} ~,~~
R_{K^*} = \frac{\text{Br}(B \to K^* \mu^+\mu^-)}{\text{Br}(B \to K^* e^+e^-)}  ~,~~
R_\phi = \frac{\text{Br}(B_s \to \phi \mu^+\mu^-)}{\text{Br}(B_s \to \phi e^+e^-)}  ~.
\end{equation}
at low $q^2$ and at high $q^2$. The SM predictions for these ratios are unity to a very high accuracy
up to kinematical effects at very low $q^2$ (cf.\ appendix~\ref{sec:pred}).
We also consider differences of $B \to K^* \ell^+ \ell^-$ angular observables as introduced in~\cite{Altmannshofer:2015mqa}\footnote{The observable $D_{P_5^\prime}$ has recently also been considered in~\cite{Capdevila:2016ivx} and~\cite{Wehle:2016yoi}, where it is referred to as $Q_5$.
See~\cite{Serra:2016ivr} for an alternative set of observables.}
\begin{eqnarray}
D_{P_5^\prime} &=& P_5^\prime(B \to K^* \mu\mu) - P_5^\prime(B \to K^* ee) ~, \\
D_{S_5} &=& S_5(B \to K^* \mu\mu) - S_5(B \to K^* ee) ~, \\
D_{A_\text{FB}} &=& A_\text{FB}(B \to K^* \mu\mu) - A_\text{FB}(B \to K^* ee) ~.
\label{eq:Dobs}
\end{eqnarray}
The angular observables $P_5^\prime$, $S_5$, and $A_\text{FB}$ do not differ significantly from their SM predictions in the high $q^2$ region across the whole NP parameter space that provides a good fit of the $b\to s \mu\mu$ data. Therefore, we consider the above LFU differences only in the low $q^2$ region. In the SM the LFU differences vanish to an excellent approximation.

\renewcommand{\arraystretch}{1.5}
\setlength\tabcolsep{0pt}
\begin{table}[tb]
\begin{center}
\begin{tabularx}{\textwidth}{ccccc}
\hline\hline
 & \multicolumn{2}{c}{(i) $C^\mu_9 - C^\mu_{10}$ fit} & \multicolumn{2}{c}{(ii) $C^\mu_9 - C^{\prime \mu}_9$ fit} \\ \hline
 \,~~~~~~~~~~~~~~~~~~~~~~~&~~~~~~~~~~ $1\sigma$ ~~~~~~~~~~&~~~~~~~~~~ $2\sigma$ ~~~~~~~~~~&~~~~~~~~~~ $1\sigma$ ~~~~~~~~~~&~~~~~~~~~~ $2\sigma$ ~~~~~~~~~~~~ \\
\hline\hline
\rowcolor[gray]{.9} $R_K^{[1,6]}$             & $0.70^{+0.09}_{-0.05}$  & $[0.59, 0.86]$   & $0.76^{+0.04}_{-0.02}$  & $[0.71, 0.84]$   \\
                    $R_K^{[15,22]}$           & $0.70^{+0.09}_{-0.05}$  & $[0.59, 0.87]$   & $0.69^{+0.05}_{-0.03}$  & $[0.64, 0.79]$   \\
\rowcolor[gray]{.9} $R_{K^*}^{[0.045,1.1]}$   & $0.87^{+0.02}_{-0.02}$  & $[0.83, 0.92]$   & $0.86^{+0.02}_{-0.01}$  & $[0.85, 0.89]$   \\
                    $R_{K^*}^{[1,6]}$         & $0.77^{+0.08}_{-0.06}$  & $[0.64, 0.92]$   & $0.76^{+0.04}_{-0.02}$  & $[0.72, 0.84]$   \\
\rowcolor[gray]{.9} $R_{K^*}^{[15,19]}$       & $0.70^{+0.09}_{-0.05}$  & $[0.59, 0.86]$   & $0.71^{+0.03}_{-0.04}$  & $[0.64, 0.79]$   \\
                    $R_{\phi}^{[1,6]}$        & $0.76^{+0.08}_{-0.06}$  & $[0.63, 0.91]$   & $0.75^{+0.04}_{-0.03}$  & $[0.70, 0.83]$   \\
\rowcolor[gray]{.9} $R_{\phi}^{[15,19]}$      & $0.70^{+0.09}_{-0.05}$  & $[0.59, 0.86]$   & $0.71^{+0.04}_{-0.05}$  & $[0.63, 0.79]$   \\
\hline\hline
                    $D_{P_5^\prime}^{[1,6]}$  & $0.29^{+0.11}_{-0.05}$  & $[0.15, 0.47]$   & $0.35^{+0.07}_{-0.07}$  & $[0.22, 0.49]$	   \\
\rowcolor[gray]{.9} $D_{S_5}^{[1,6]}$         & $0.12^{+0.05}_{-0.02}$  & $[0.06, 0.2]$    & $0.15^{+0.03}_{-0.03}$  & $[0.09, 0.21]$   \\
                    $D_{A_\text{FB}}^{[1,6]}$ &$-0.09^{+0.02}_{-0.02}$  &$[-0.13, -0.04]$  &$-0.10^{+0.02}_{-0.02}$  &$[-0.14, -0.06]$ \\
\hline\hline
\end{tabularx}
\end{center}
\caption{Predictions for lepton flavour universality ratios and differences in new physics models with muon specific contributions to $C_9$ and $C_{10}$, or $C_9$ and $C_9^\prime$. The superscripts on the observables indicate the $q^2$ range in GeV$^2$. }
\label{tab:LFU}
\end{table}

\begin{figure}[tb]
\centering
\includegraphics[width=0.48\textwidth]{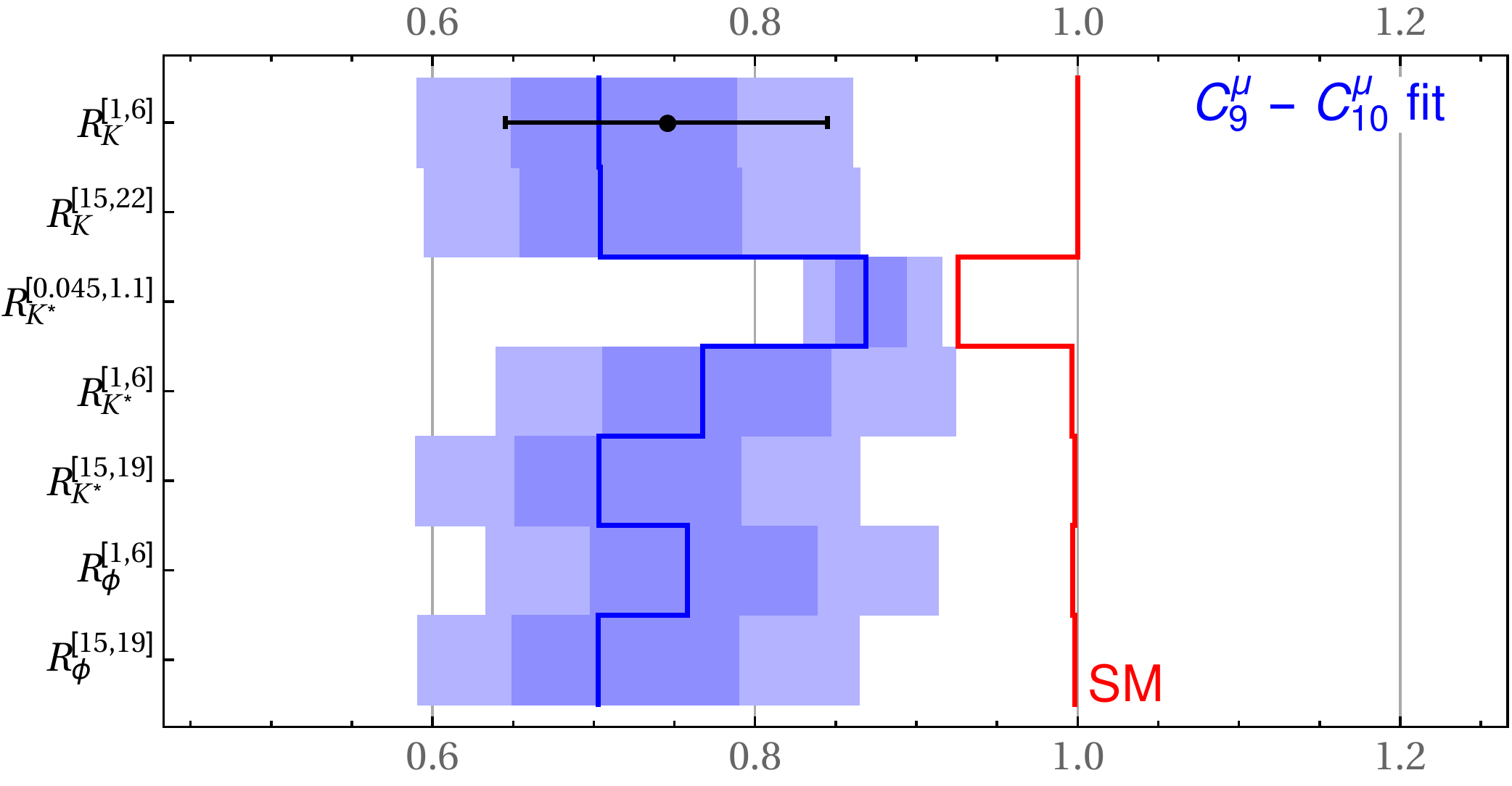} ~~~
\includegraphics[width=0.48\textwidth]{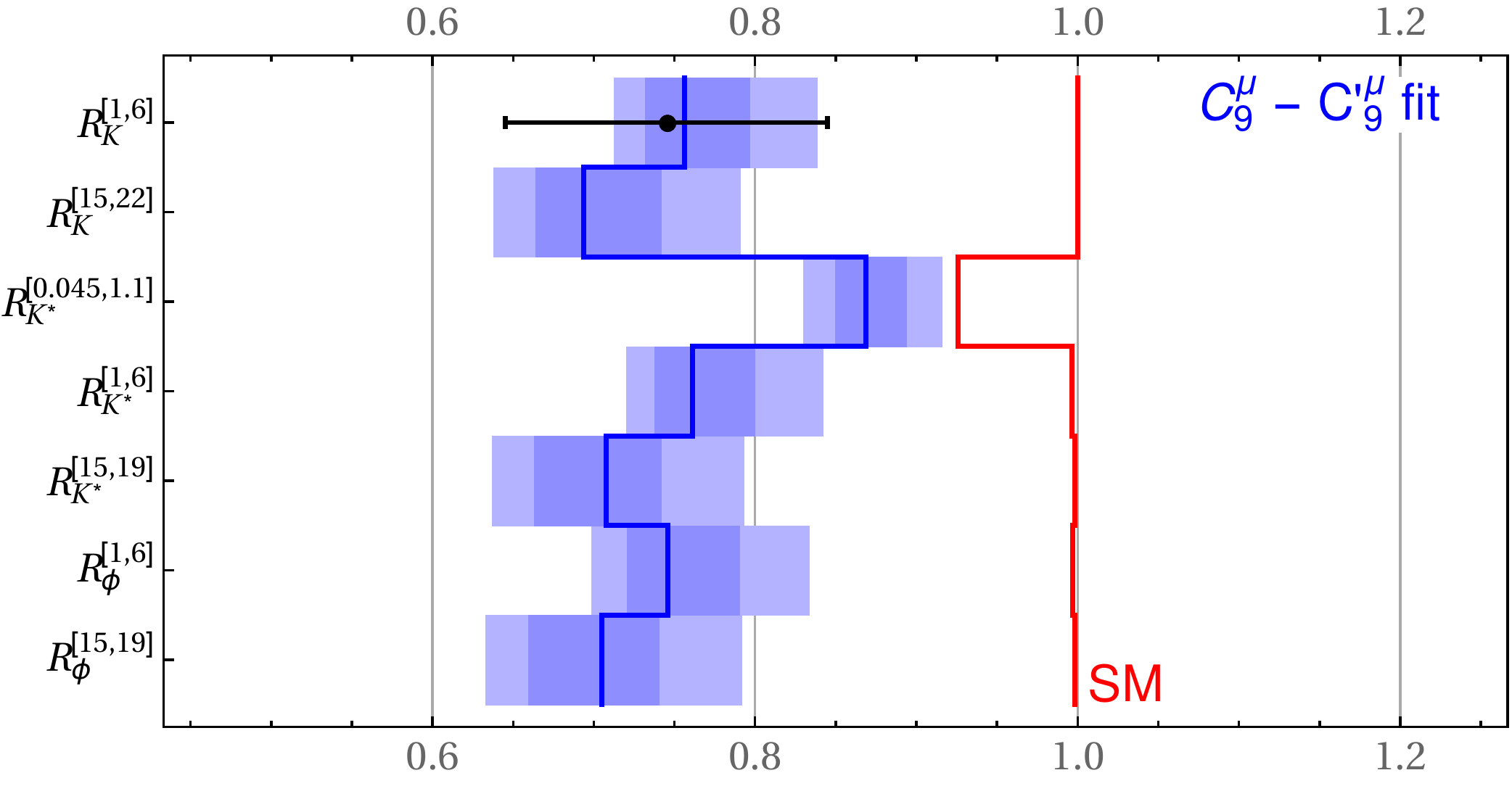} \\[12pt]
\includegraphics[width=0.48\textwidth]{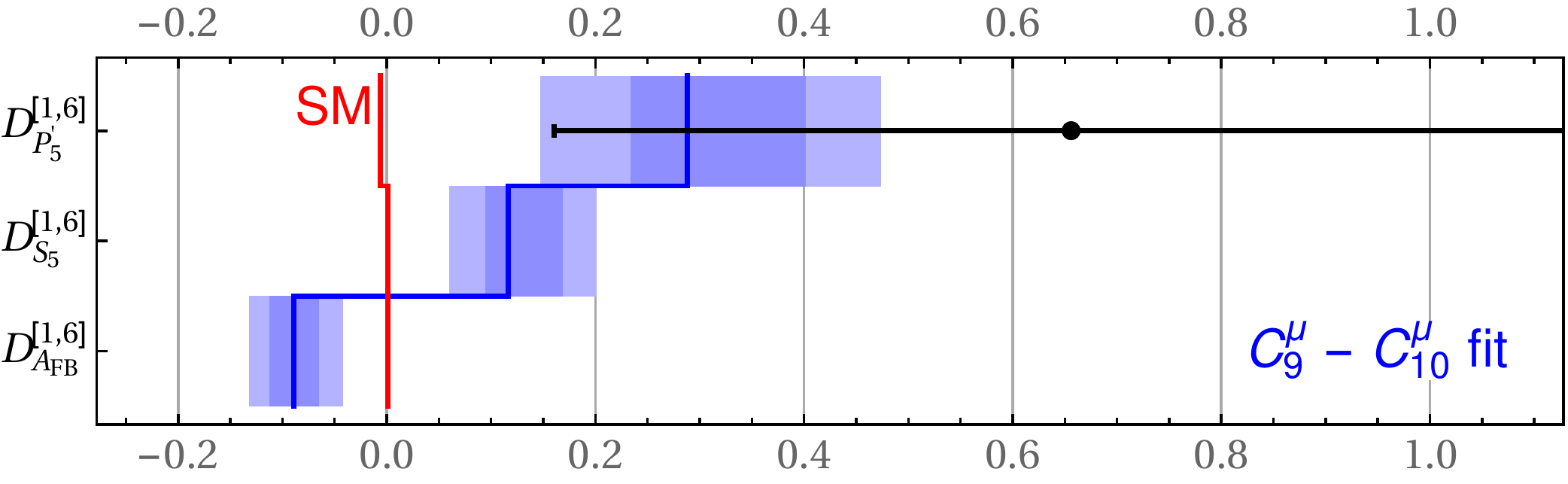} ~~~
\includegraphics[width=0.48\textwidth]{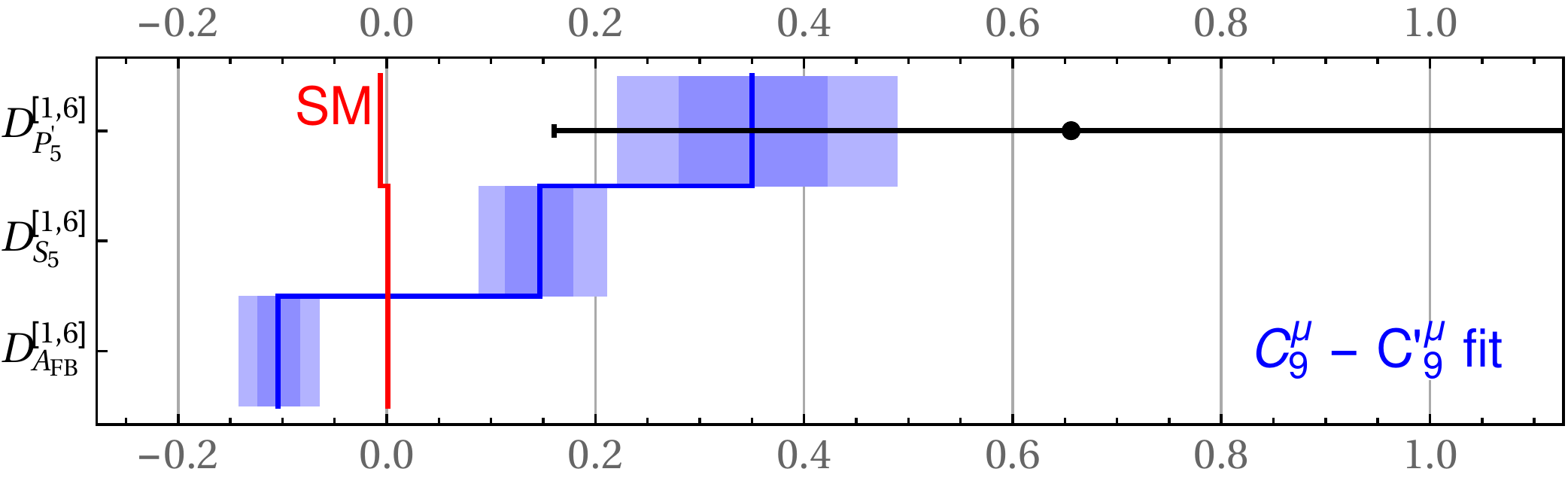} \\[12pt]
\caption{Predictions for lepton flavour universality ratios and differences in new physics models with muon specific contributions to $C_9$ and $C_{10}$, or $C_9$ and $C_9^\prime$. The superscripts on the observables indicate the $q^2$ range in GeV$^2$. The red lines show the SM predictions. The $1\sigma$ and $2\sigma$ ranges in the NP scenarios are shown in blue. In black the LHCb measurement of $R_K$ and the Belle measurement of $D_{P_5^\prime}$. }
\label{fig:LFU}
\end{figure}

In Tab.~\ref{tab:LFU} and in Fig.~\ref{fig:LFU} we show the predictions for the LFU observables for two scenarios: (i) new physics in the Wilson coefficients $C_9$ and $C_{10}$; (ii) new physics in the Wilson coefficients $C_9$ and $C_9^\prime$.
We observe that in both scenarios, the observables $R_K$, $R_{K^*}$ and $R_\phi$ are all suppressed with respect to their SM predictions.
Since the best-fit regions of both scenarios correspond to similar values of the Wilson coefficients -- a sizable shift in $C_9^\mu$ and
small effects in $C_{10}^\mu$ or $C_9^{\prime\mu}$, respectively -- the predictions for the observables are very similar both for
the branching ratios and for the angular observables.
The LHCb measurement of $R_K$~\cite{Aaij:2014ora} is in excellent agreement with our predictions.
The recent results on $D_{P_5^\prime}$ by Belle~\cite{Wehle:2016yoi} are compatible with our predictions but still afflicted by large statistical uncertainties.
If future measurements of any of the discussed LFU observables shows significant discrepancy with respect to SM predictions, it would be clear evidence for new physics.

\section{Conclusions}

In this paper, we have analyzed the status of the ``$B\to K^*\mu^+\mu^-$
anomaly'', i.e.\ the tension with SM predictions in various $b\to s\mu^+\mu^-$
processes, after the new measurements of $B\to K^*\mu^+\mu^-$ angular observables
by ATLAS and CMS and including updated measurements by LHCb.
We find that the significance of the tension remains strong.
Assuming the tension to be due to NP, a good fit is obtained with a negative NP
contribution to the Wilson coefficient $C_9$. Models predicting the NP contributions
to the coefficients $C_9$ and $C_{10}$ to be equal with an opposite sign give
a comparably good fit.

We also studied the $q^2$ and helicity dependence of the non-standard contribution to $C_9$. We find that the data agrees well with a $q^2$ and helicity independent new physics effect in $C_9$. A hadronic effect with these properties might appear surprising, but cannot be excluded as an explanation of the tensions.

Finally, again under the hypothesis of NP explaining the tensions, we provided a set of predictions for LFU observables. Assuming that the new physics affects only $b \to s \mu \mu$ but not $b \to s e e$ transitions, we confirm that the latest $B \to K^* \mu^+\mu^-$ data shows astonishing compatibility with the LHCb measurement of the LFU ratio $R_K$.
Future measurements of LFU observables that show significant deviations from SM predictions could not be explained by underestimated hadronic contributions but would be clear evidence for a new physics effect.

\section*{Acknowledgments}

We thank Ayan Paul, Javier Virto, Jure Zupan, and Roman Zwicky for useful comments.
WA acknowledges financial support by the University of Cincinnati.
DS thanks
Christoph Langenbruch for reporting a bug in flavio and
the organizers of the LHCb Workshop in Neckarzimmern for hospitality
while this paper was written.
The work of CN, PS, and DS was supported by the DFG cluster of
excellence ``Origin and Structure of the Universe''.

\appendix

\section{Predictions}\label{sec:pred}

\begin{figure}[tbp]
\centering
\includegraphics[width=\textwidth]{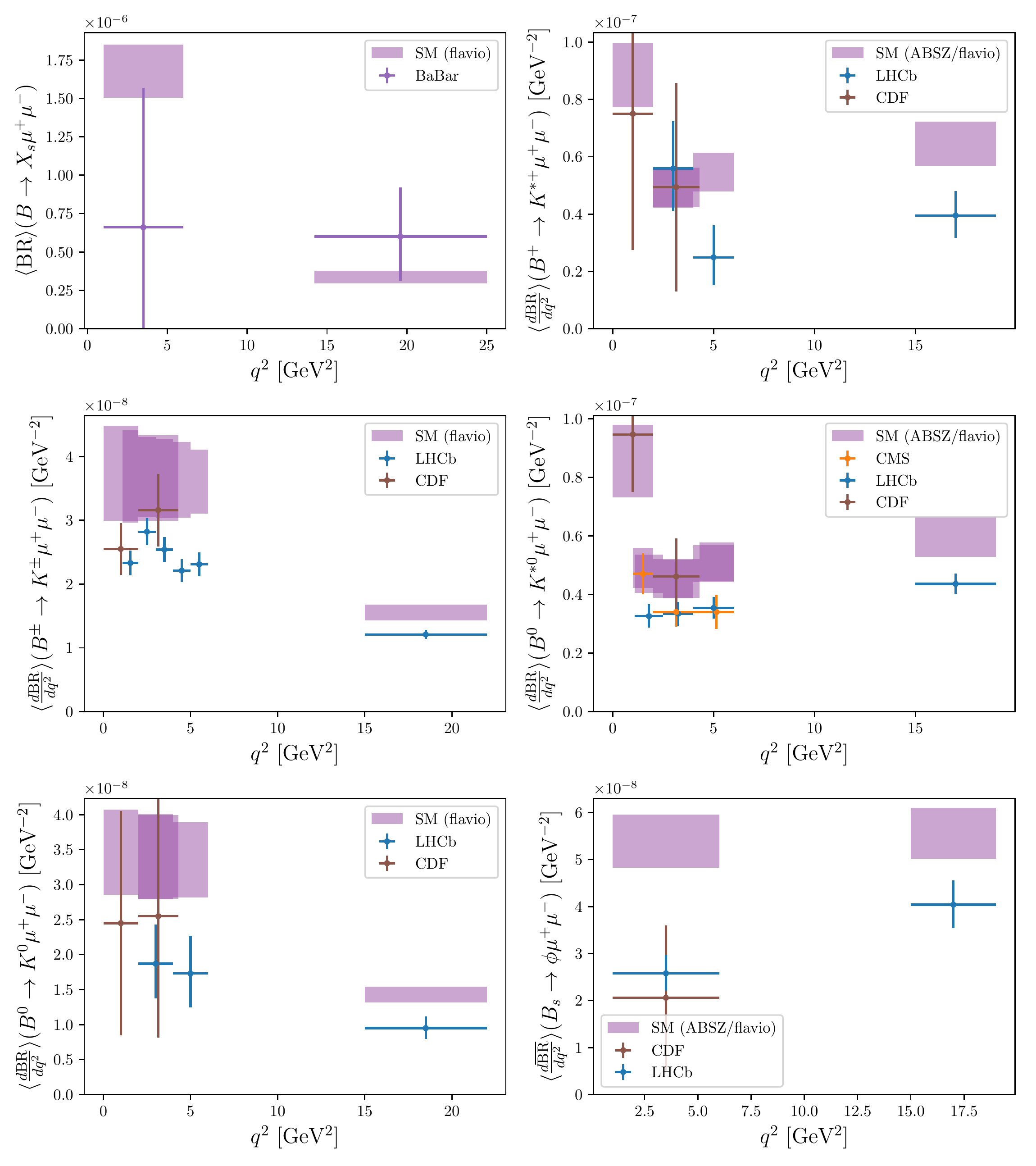}
\caption{Experimental measurements vs.\ SM predictions for the branching ratios.
``ABSZ'' refers to~\cite{Altmannshofer:2014rta,Straub:2015ica}.}
\label{fig:pred:br}
\end{figure}

\begin{figure}[tbp]
\centering
\includegraphics[width=\textwidth]{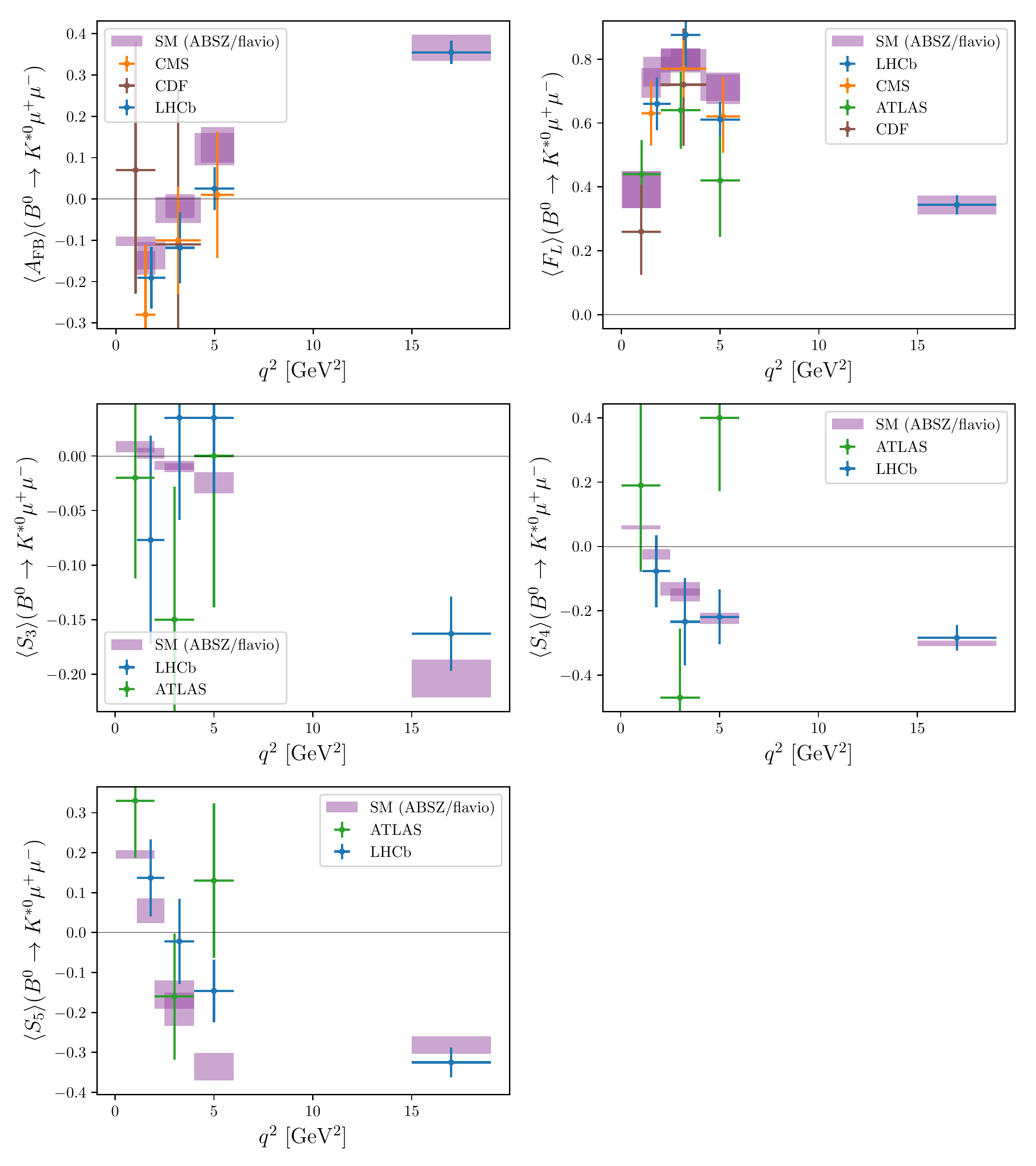}
\caption{Experimental measurements vs.\ SM predictions for the $B\to  K^*\mu^+\mu^-$ angular observables.
``ABSZ'' refers to~\cite{Altmannshofer:2014rta,Straub:2015ica}.}
\label{fig:pred:ks-s}
\end{figure}

\begin{figure}[tbp]
\centering
\includegraphics[width=\textwidth]{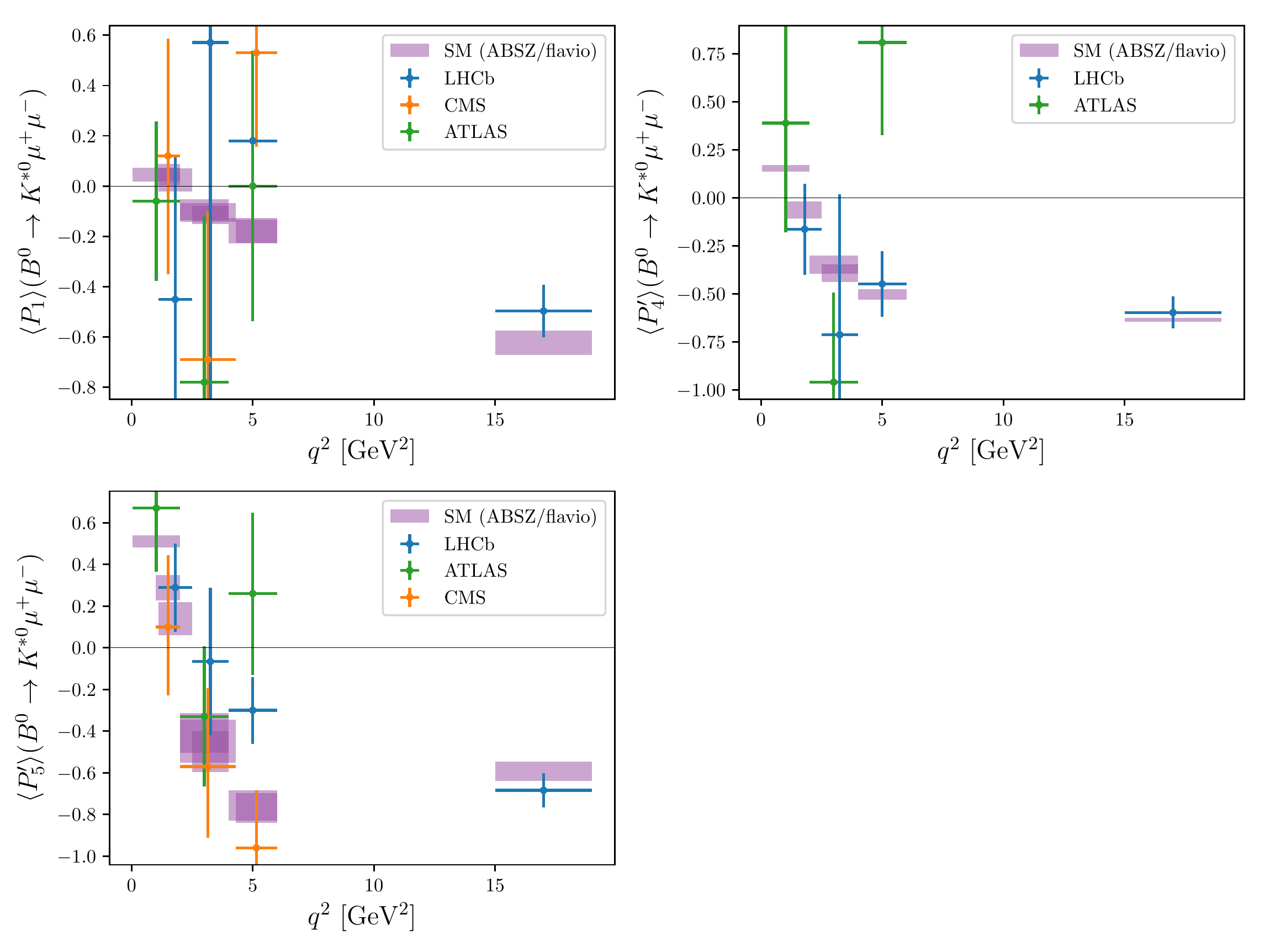}
\caption{Experimental measurements vs.\ SM predictions for the $B\to  K^*\mu^+\mu^-$ ``optimized'' observables.
``ABSZ'' refers to~\cite{Altmannshofer:2014rta,Straub:2015ica}.}
\label{fig:pred:ks-p}
\end{figure}

\begin{figure}[tbp]
\centering
\includegraphics[width=\textwidth]{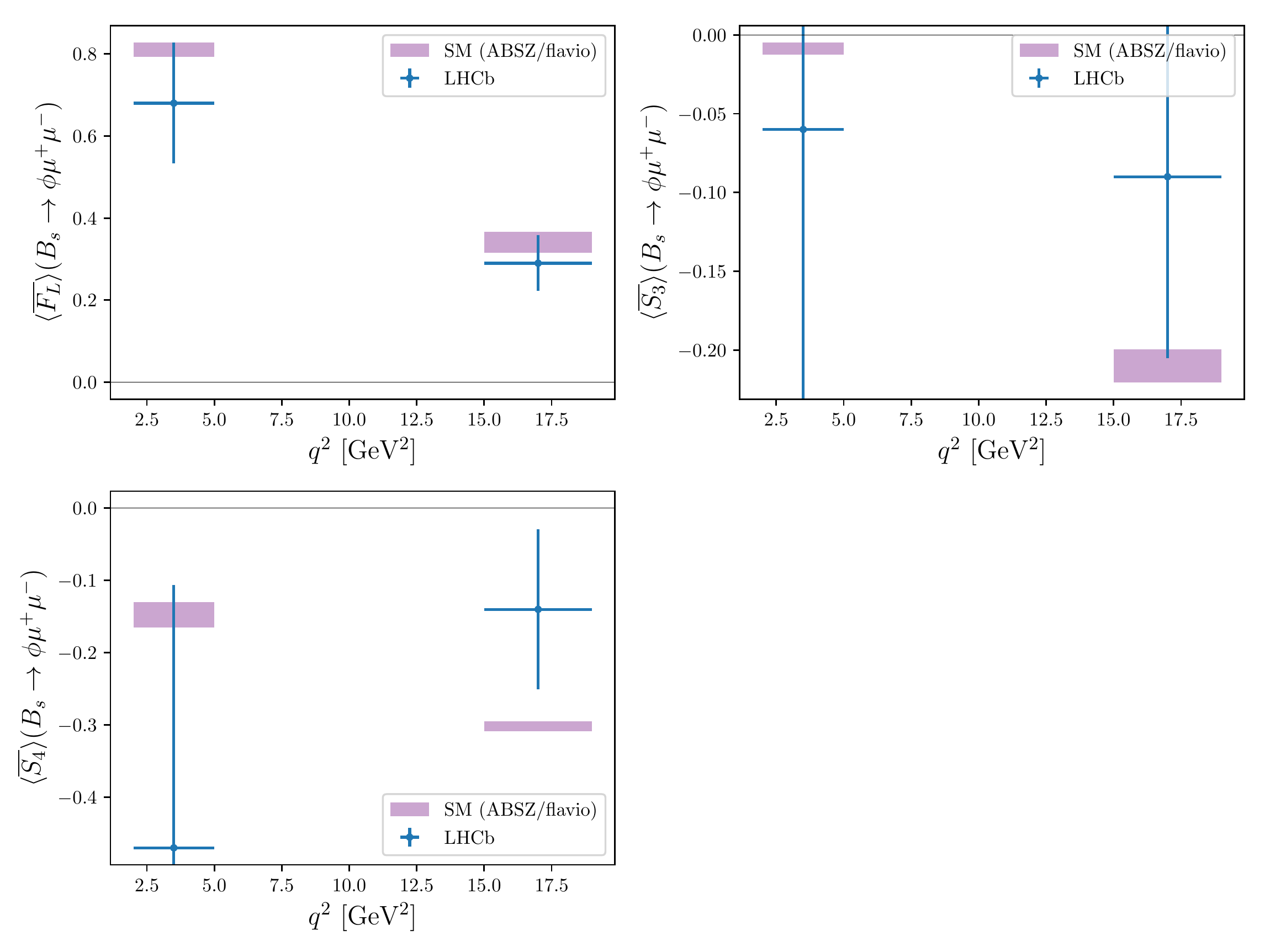}
\caption{Experimental measurements vs.\ SM predictions for the $B_s\to  \phi\mu^+\mu^-$ angular observables.
``ABSZ'' refers to~\cite{Altmannshofer:2014rta,Straub:2015ica}.}
\label{fig:pred:phi-s}
\end{figure}

Figures~\ref{fig:pred:br}--\ref{fig:pred:phi-s} compare the binned experimental
measurements to the SM predictions in the same bins, obtained with \texttt{flavio}
version~0.21.2.
We only show the bins included in our fits (cf.\ the discussion in section~\ref{sec:Heff}).
``ABSZ'' refers to the predictions for $B\to  V\ell^+\ell^-$
observables in \texttt{flavio} which are based on the results of
\cite{Straub:2015ica} (BSZ) for low $q^2$ and~\cite{Altmannshofer:2014rta} (AS)
for high $q^2$.

\begin{table}
\begin{center}
\begin{tabularx}{\textwidth}{cccc}
\hline\hline
~~~~~~~~~~~~ & ~~~~~~~~ $q^2 \in [0.045, 1.1]$ ~~~~~~~~ & ~~~~~~~~ $q^2 \in [1.1, 6.0]$ ~~~~~~~~ & ~~~~~~~~ $q^2 \in [15.0, 19.0]$ ~~~~~~~~ \\
\hline
\rowcolor[gray]{.9} $\mathcal{R}_{K^*}$ &  $0.9259(41)$  & $0.9965(6)$  & $0.9981(1)$ \\
                    $\mathcal{R}_{\phi}$&  $0.9299(28)$  & $0.9970(2)$  & $0.9981(1)$ \\
\rowcolor[gray]{.9} $D_{P_5^\prime}$    &  $0.0936(37)$ & $-0.0064(5)$ & $-0.0008(1)$ \\
                    $D_{S_5}$           & $-0.0402(26)$  & $0.0008(4)$  & $0.00022(4)$ \\
\rowcolor[gray]{.9} $D_{A_\text{FB}}$   &  $0.0088(5)$   & $0.0008(3)$ & $-0.00028(5)$ \\
\hline\hline
\end{tabularx}
\caption{SM predictions for LFU observables in different $q^2$ bins. The $D$
observables have been defined in eq.~\eqref{eq:Dobs}.}
\label{tab:LFUpredictions}
\end{center}
\end{table}
Table \ref{tab:LFUpredictions} shows the SM predictions for observables
sensitive to violation of LFU. The uncertainties are parametric uncertainties
only, i.e.\ it is assumed that final state radiation effects are simulated fully
on the experimental side
and QED corrections due to light hadrons are neglected
(cf.~\cite{Bordone:2016gaq}).

\clearpage

\bibliographystyle{JHEP}
\bibliography{bibliography}

\end{document}